\documentclass[10pt]{iopart}
\usepackage[utf8]{inputenc}
\usepackage{iopams}
\usepackage{graphicx}
\usepackage{cite}
\newcommand{\text}[1]{\mathrm{#1}}

\begin{document}

\title[Field enhancement due to ion motion]{Accelerating field enhancement due to ion motion in plasma wakefield accelerators}

\author{V.A. Minakov}
\address{Budker Institute of Nuclear Physics, Novosibirsk, 630090, Russia}
\address{Novosibirsk State University, Novosibirsk, 630090, Russia}
\ead{V.A.Minakov@inp.nsk.su}
\author{A.P. Sosedkin}
\address{Budker Institute of Nuclear Physics, Novosibirsk, 630090, Russia}
\address{Novosibirsk State University, Novosibirsk, 630090, Russia}
\ead{A.P.Sosedkin@inp.nsk.su}
\author{K.V. Lotov}
\address{Budker Institute of Nuclear Physics, Novosibirsk, 630090, Russia}
\address{Novosibirsk State University, Novosibirsk, 630090, Russia}
\ead{K.V.Lotov@inp.nsk.su}

\vspace{10pt}
\begin{indented}
\item[]\today
\end{indented}

\begin{abstract}
Ion motion in plasma wakefield accelerators can cause temporal increase of the longitudinal electric field shortly before the wave breaks. The increase is caused by  re-distribution of the wave energy in transverse direction and may be important for correct interpretation of experimental results and acceleration of high-quality beams.
\end{abstract}

\vspace{2pc}
\ioptwocol

\section{Introduction}

Controllable electric fields in plasmas can exceed those in radio-frequency accelerating structures by orders of magnitude, and this ability promises size reduction of future particle accelerators. Of particular interest in this context is wakefield acceleration \cite{NatPhot7-775,RAST9-19,RAST9-63,RAST9-85,RMP90-035002}, in which the field is excited in the plasma for a short time between passage of the drive beam (laser pulse or particle bunch) and the accelerated witness beam.

The time interval between driver and witness is usually one or two periods of Langmuir wave. It is too short for heavy plasma ions to respond, so only plasma electrons participate in wave formation, while ions remain immobile. However, there are cases in which ion motion becomes important. These are very intense drivers or witnesses that perturb the ion background during a single wave period \cite{PRL95-195002,PRL104-155001,PRL118-244801,PRAB20-111301,PRL121-264802} and long drivers that resonantly excite the plasma wave during many wave periods \cite{PPR28-125,PRL109-145005,PoP21-056705,JPB47-234003}. Our study refers to the second case.

The effect of ion motion on the plasma wave is usually destructive. Ion density perturbations cause phase mismatch of electron oscillations and wavebreaking \cite{PRL86-3332,PoP10-1124,PRL109-145005,PoP21-056705,PoP25-103103}. However, at certain conditions, the ion motion results in increase of the longitudinal electric field near the axis. The increase is not drastic, about 20\%, but this value may be important for correct interpretation of experimental results and acceleration of high-quality witnesses. We describe the discovered effect in Sec.\,\ref{s2} and suggest its explanation in Sec.\,\ref{s3}.

\section{Wakefield enhancement}\label{s2}

The effect of wakefield enhancement was first noticed in numerical simulations of the AWAKE experiment. In AWAKE \cite{NIMA-829-76,PPCF60-014046,Nat.561-363}, a long proton bunch undergoes seeded self-modulation in the plasma \cite{PRL104-255003,PoP18-103101,PoP22-103110} and transforms into a train of short micro-binches that resonantly drive the plasma wave \cite{PRL122-054801,PRL122-054802}. The number of micro-bunches depends on the plasma density and is typically about one hundred, so the plasma wave exists sufficiently long to move the ions. For this reason, AWAKE simulations usually include ion motion.

\begin{figure}[tb]\centering
 \includegraphics{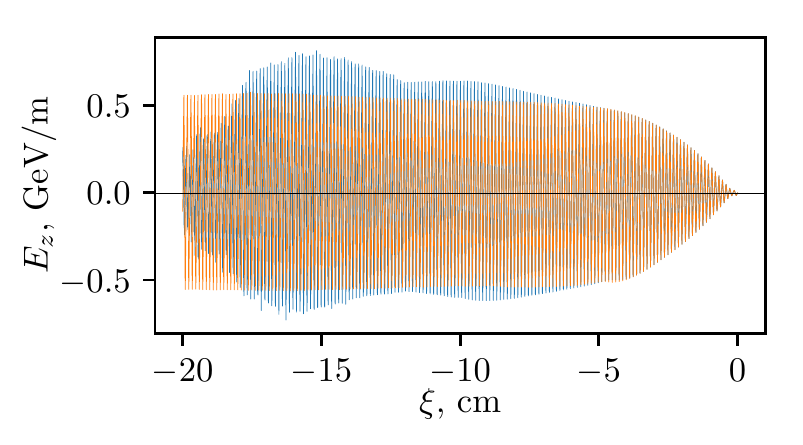}
\caption{Dependence of the on-axis electric field $E_z$ on the co-moving coordinate $\xi$ at $z=z_0$ for immobile (orange) and mobile (blue) ions.}
\label{fig_ez}
\end{figure}

% Проверить и поправить параметры
\begin{table}[tb]
 \caption{Parameters of the illustration variant.}\label{t1}
 \begin{center}\begin{tabular}{ll}\hline
  Parameter \& notation & Value \\ \hline
  Plasma density, $n_0$ & $7\times 10^{14}\,\text{cm}^{-3}$ \\
  Distance to observation point, $z_0$ & 5\,m \\
  Plasma ion-to-electron mass ratio & 157\,000\\
  Maximum beam density, $n_{b0}$ & $6.9\times 10^{12}\,\text{cm}^{-3}$ \\
  Beam half-length, $\sigma_{zb}$ & 7\,cm  \\
  Beam radius, $\sigma_{rb}$ & 0.2\,mm  \\
  Beam energy, $W_b$ & 400\,GeV  \\
  Beam energy spread, $\delta W_b$ & 0.35\%  \\
  Beam normalized emittance, $\epsilon_{bn}$, & $3$ mm mrad \\
  \hline
 \end{tabular}\end{center}
\end{table}

In simulations of various AWAKE regimes, the longitudinal electric field $E_z$ on the axis reaches higher values if ions are allowed to move (Fig.\,12 in \cite{NIMA-829-3}). Figure~\ref{fig_ez} illustrates the effect for plasma and beam parameters listed in Table~\ref{t1}. The proton beam density before entering the plasma at $z=0$ is
\begin{eqnarray}\nonumber
 \fl n_b = \frac{n_{b0}}{2} \, e^{-r^2/2 \sigma_{rb}^2} & \left[  1 + \cos \left( \sqrt{\frac{\pi}{2}} \frac{\xi}{\sigma_{zb}}  \right)  \right], \\
\label{e1}
  & -\sigma_{zb} \sqrt{2\pi} < \xi < 0,
\end{eqnarray}
and zero otherwise. Here $\xi = z-ct$ is the co-moving coordinate, and $c$ is the speed of light. The half-cut cosine profile (\ref{e1}) mimics plasma creation with rapid ionization of a neutral gas by a short laser pulse that co-propagates with the beam centroid. Simulations are made with quasi-static two-dimensional (axisymmetric) code LCODE \cite{PRST-AB6-061301,NIMA-829-350}.

As we see in Fig.\,\ref{fig_ez}, the field enhancement is strongest shortly before wavebreaking, after which the longitudinal field quickly degrades in case of mobile ions. The effect cannot come from narrow field spikes, as the wakefield potential (in which all local spikes are averaged out) reveals the same trend.

\begin{figure}[htb]\centering
 \includegraphics{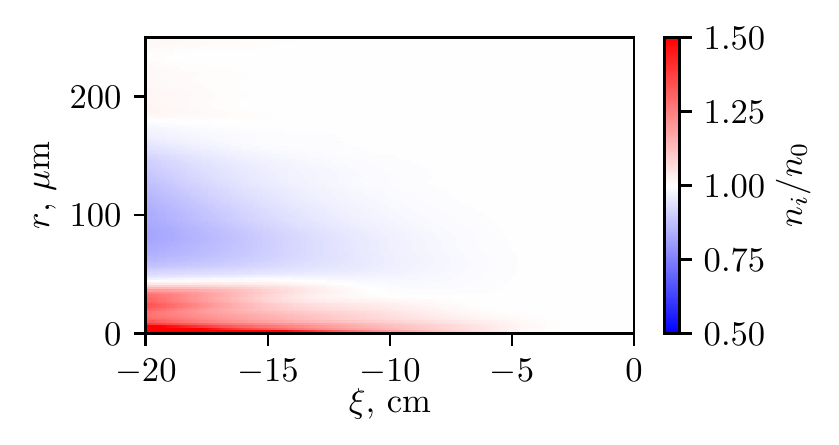}
\caption{Ion density perturbation at $z=z_0$.}
\label{fig_ni}
\end{figure}
% Исправить на -S_z
% Чуть понизить (a), (b), ... (как на рис.4)
\begin{figure*}[tb]\centering
 \includegraphics{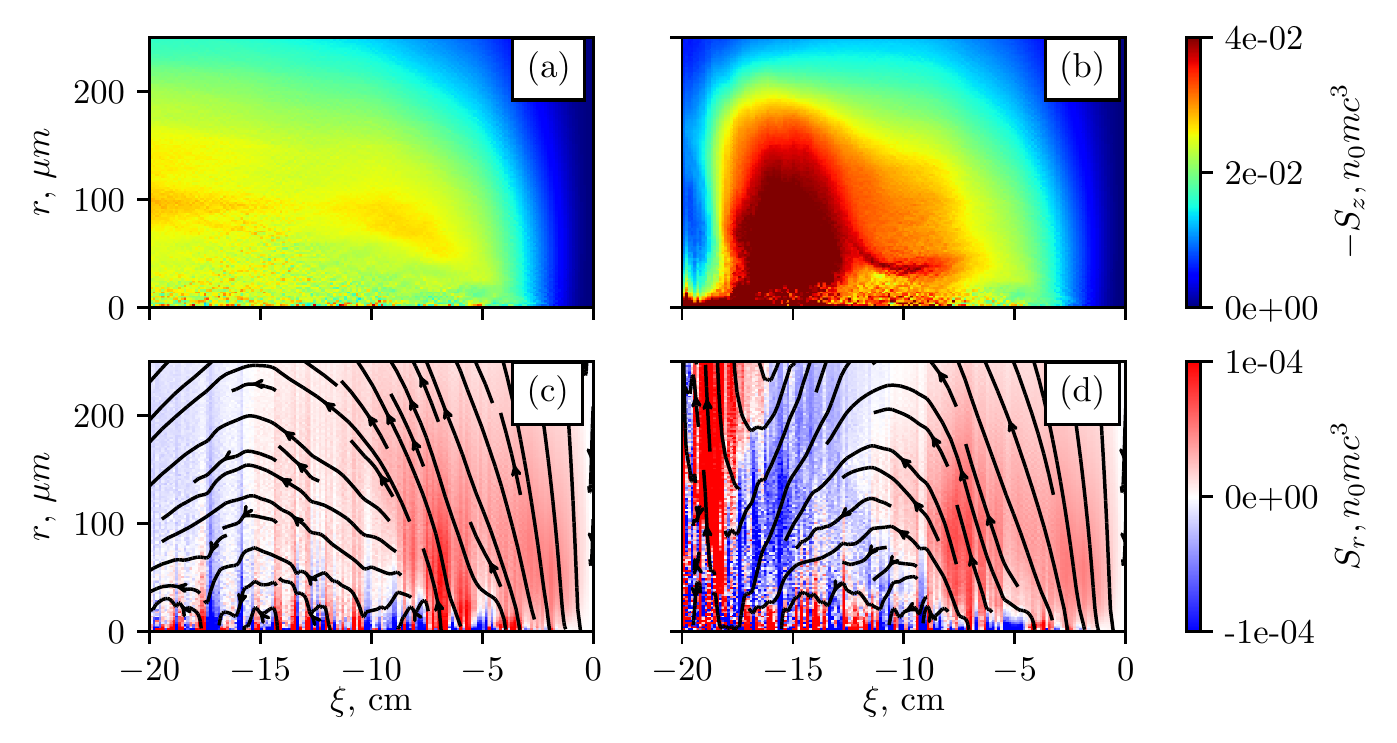}
\caption{Longitudinal (a,b) and radial (c,d) components of the energy flux density $\vec{S}$ in the co-moving window for immobile (a,c) and mobile (b,d) ions. Black arrows show directions of energy flow.}
\label{fig_s}
\end{figure*}

\section{Cause of wakefield enhancement}\label{s3}

The onset of wakefield enhancement obviously correlates with visible perturbations of the ion density $n_i$ (Fig.\,\ref{fig_ni}). The density increases near the axis and decreases off-axis, which is typical for wakefields of narrow drivers \cite{PRL86-3332,PoP10-1124,PRL109-145005,PoP25-103103}. This density profile is known to distort wave fronts, which causes energy transfer from longitudinal to transverse oscillations \cite{PoP10-1124}. Therefore, the increase of wave front curvature cannot account for the growth of the longitudinal field.

The effect comes from spatial redistribution of the wave energy. To make it visible, we look at the energy flux density in the co-moving frame \cite{PRE69-046405}
\begin{eqnarray}\nonumber
 \fl \vec{S} = \sum_i (\gamma_i - 1) m_i c^2 (\vec{v}_i - c \vec{e}_z) & - c \vec{e}_z \frac{E^2+B^2}{8\pi} \\
\label{e2}
 & + \frac{c}{4 \pi} \left[ \vec{E} \times \vec{B}  \right],
\end{eqnarray}
where $m_i$, $\vec{v}_i$, and $\gamma_i$ are mass, velocity, and relativistic factor of plasma particles, the summation is carried out over plasma particles in the unit volume, $\vec{E}$ and $\vec{B}$ are electric and magnetic fields, and $\vec{e}_z$ is the unit vector in $z$-direction. The energy in the co-moving window predominantly flows against $z$-direction, so the component $S_z$ shows where the wakefield energy is (Fig.\,\ref{fig_s}). Unlike the usual energy density, the energy flux density $S_z$ does not oscillate with the plasma frequency \cite{PoP25-103103}, which eases visualisation. 

\begin{figure*}[htb]\centering
 \includegraphics{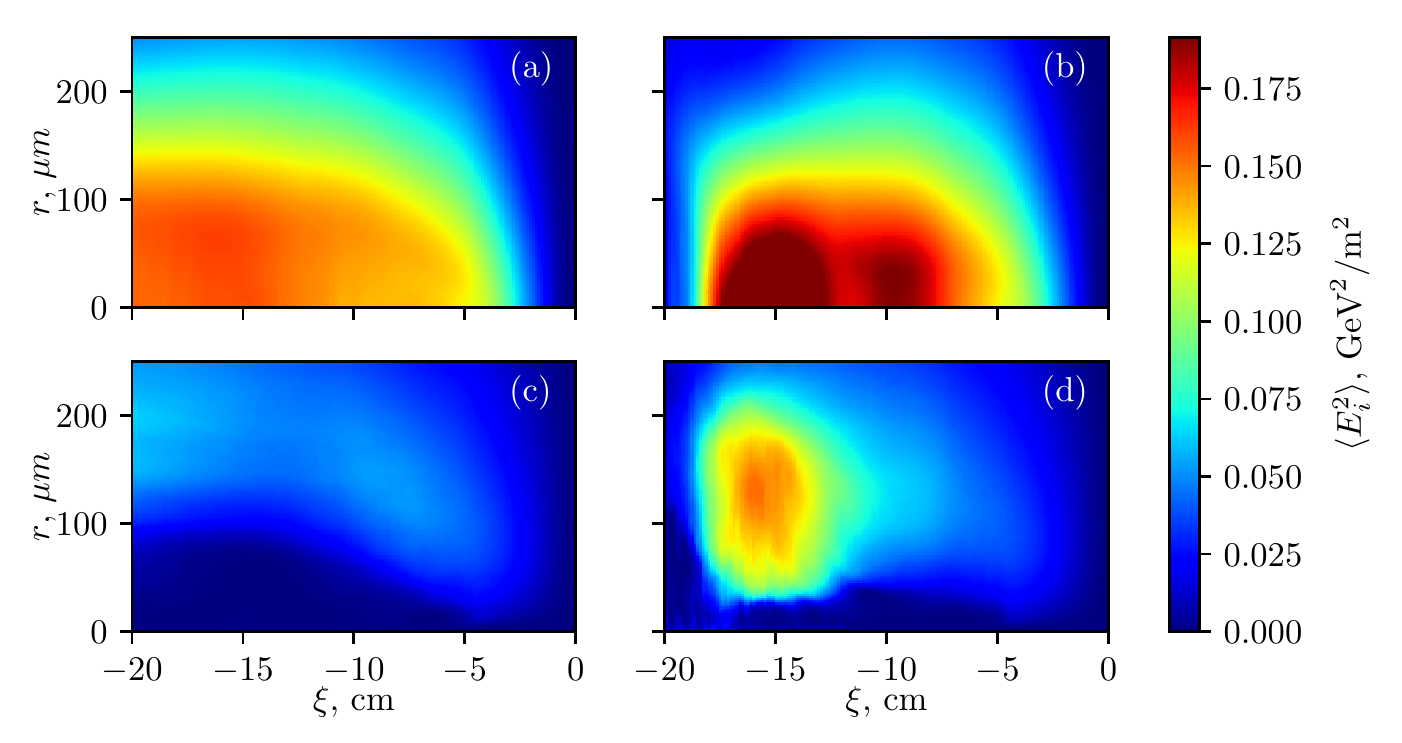}
\caption{Longitudinal (a,b) and radial (c,d) components of the electric field for immobile (a,c) and mobile (b,d) ions. As the field oscillates with the plasma frequency, we show the period-averaged squares $\langle E_z^2 \rangle$ and $\langle E_z^2 \rangle$.}
\label{fig_e}
\end{figure*}

In the case of immobile ions, the energy remains at almost the same radial positions where it was delivered to the plasma by the beam [Fig.\,\ref{fig_s}(a,c)], which is typical for weakly nonlinear waves \cite{PRE69-046405}. If ions are allowed to move, there appears energy redistribution along the radius [Fig.\,\ref{fig_s}(b,d)]. This is indeed the energy redistribution, as there is almost no beam in this region ($|\xi| \gtrsim 14\,\text{cm} = 2\sigma_{zb}$), and the total energy flux
\begin{equation}\label{e3}
    \Psi (\xi) = \int_0^\infty S_z 2 \pi r \, dr
\end{equation}
is conserved. The energy concentrates near the axis and amplifies the wakefield amplitude there (Fig.\,\ref{fig_e}). The effect dominates over energy transfer to the transverse field component [which certainly takes place, Fig.\,\ref{fig_e}(d)], and we observe the overall enhancement of the longitudinal field [Fig.\,\ref{fig_e}(c)]. Similar redistribution of energy may be responsible for temporal increase of the on-axis electric field immediately after wave-breaking in the case of laser-driven wakefield (Fig.1(b) of \cite{PoP25-103103}).

% Подпись: $\langle E_i^2 \rangle$,~$\text{GeV}^2/\text{m}^2$

Figure~\ref{fig_s} contains another interesting feature not related to amplitude enhancement. After the wave breaks at $|\xi| \approx 18\,$cm, the longitudinal energy flux disappears [Fig.\,\ref{fig_s}(b)], because the wake energy goes away in radial direction with high-energy electrons that appear after wave-breaking [Fig.\,\ref{fig_s}(d)].

The question, however, remains why the wave energy drifts radially on this ion background. We have no answer yet. Apparently, this is not simply related to the ion density gradient, as gradients of both signs are present in the system, and we do not observe energy drift in different directions.

%\ack
% Пока что эта работа планируется в отчет по госзаданию. Там благодарность не требуется.
%This work is supported by The Russian Science Foundation, grant No.~????. 
%The computer simulations are made at Siberian Supercomputer Center SB RAS (???).

\end{document}